\documentclass[journal,final,10pt]{IEEEtran}
\usepackage[T1]{fontenc}

\usepackage{cite}

\ifCLASSINFOpdf
   \usepackage[pdftex]{graphicx}
  \DeclareGraphicsExtensions{.pdf}
\else
  \usepackage[dvips]{graphicx}
  \DeclareGraphicsExtensions{.eps}
\fi

\usepackage{amsmath, mathtools}
\usepackage[cmintegrals]{newtxmath}
\usepackage{color}

{}

\newtheorem{proposition}{Proposition}
\newtheorem{corol}{Corollary}
\DeclareMathOperator{\tr}{tr}

\usepackage{paralist}
\usepackage{enumerate}
\usepackage{enumitem}
\setlist[enumerate]{noitemsep, topsep=0pt}

\usepackage{array}

\ifCLASSOPTIONcompsoc
    \usepackage[caption=false,font=normalsize,labelfont=sf,textfont=sf]{subfig}
\else
    \usepackage[caption=false,font=footnotesize]{subfig}
\fi

\usepackage{url}

\begin{document}
%
\title{On the Achievable Rate of Bi-Static Modulated Re-Scatter Systems}

\author{Ruifeng~Duan,~\IEEEmembership{Member,~IEEE,}
        Riku~J\"{a}ntti,~\IEEEmembership{Senior~Member,~IEEE,}
        H\"{u}seyin~Yi\u{g}itler and 
        Kalle~Ruttik,~\IEEEmembership{Member,~IEEE}
\thanks{Copyright (c) 2015 IEEE. Personal use of this material is permitted. However, permission to use this material for any other purposes must be obtained from the IEEE by sending a request to pubs-permissions@ieee.org.}
\thanks{This work was supported in part by the Academy of Finland under Grant no. 311760, and by Nokia Bell Labs under a donation.}
\thanks{The authors are with the Department of Communications and Networking, Aalto University, Espoo, 02150 Finland. (e-mail: \{firstname.surname\}@aalto.fi).}
}

%
%

\markboth{Accepted for publication}{}%

\maketitle

\begin{abstract}
In ambient re-scatter\footnote{Widely used term \emph{backscatter} refers to reflecting the received signals back in the direction of arrival. In this work, the scattering direction is not constrained so that the term \emph{re-scatter} is used instead.} communications, devices convey information by modulating and re-scattering the radio frequency signals impinging on their antennas.  In this correspondence, we consider a system consisting of a legacy modulated continuous carrier multiple-input-multiple-output (MIMO) link and a multi-antenna modulated re-scatter (MRS) node, where the MRS node modulates and re-scatters the signal generated by the legacy transmitter. The receiver seeks to decode both the original message and the information added by the MRS. We show that the achievable sum rate of this system exceeds that which the legacy system could achieve alone. We further consider the impact of channel estimation errors under the least squares channel estimation and study the achievable rate of the legacy and MRS systems, where a linear minimum mean square error receiver with successive interference cancellation is utilized for joint decoding.
\end{abstract}


%
\IEEEpeerreviewmaketitle

\section{Introduction}
\IEEEPARstart{O}{ne} of the limiting factors of connecting things to the Internet using wireless technologies is the availability of energy. One solution is provided by the modulated back-scattering (MBS) systems, e.g. radio-frequency identification, where the tags modulate their information onto a carrier generated by a reader, and the reader decodes the modulated information. In the advanced forms of MBS systems, multiple-antenna techniques are applied to increase the achievable rate and improve the reliability \cite{Boyer2014, He2015a}. Their communication range can be increased by using a bi-static setup, in which the transmitter and receiver are physically separated \cite{Kimionis2014}. Recently, a new communication technology, referred to as an ambient MBS system, is emerging where the tags can backscatter ambient modulated signals with added information \cite{Bharadia2015}. Although the circuit power consumption could still be a serious problem in practice \cite{Wu2016,Wu2016a}, it is nevertheless much smaller than in case of active transmitters \cite{Bharadia2015}. The symbol detection and bit error rate have been deduced for such a system \cite{Wang2016, Qian2017}. In addition, the performance of such a system in the single antenna case has been investigated in \cite{Darsena2017}.

To the best of our knowledge, however, the previous models did not consider joint decoding of the legacy and backscatter systems.These systems enhance the use-case possibilities of MBS systems since the information is exchanged through modulating and back-scattering the radio frequency signals without the need of having power-hungry transceivers.

In this paper, we\begin{inparaenum}[ 1)~]
\item{propose a new system that extends a bi-static MBS system to an ambient bi-static modulated re-scatter (MRS) system, which allows information transmission between a multi-antenna transmitter and a multi-antenna receiver (legacy MIMO system);} \item{show that, with full channel state information at the receiver (CSIR), the achievable sum rate (ASR) of the proposed system exceeds that which the legacy MIMO system could achieve alone;}\item{study the asymptotic limiting achievable rate (LAR) as either the number of transmit antennas or the number of receive antennas approaches infinity;} \item{propose a simple pilot structure using Hadamard matrix for channel estimation, which enables the receiver to jointly estimate the direct and re-scattered paths;} \item{consider the channel estimation errors of the least squares (LS) estimator and provide the associated achievable rate lower bound.}
\end{inparaenum}

In this design, the multi-antenna MRS node bears additional information on the signals emitted by the legacy transmitter, and the receiver decodes the information of both sources as shown in Fig.~\ref{Fig:Channel}. The excess rate can be achieved by the legacy MIMO system alone or it can be shared between the two systems. The LAR of the considered system is the LAR of a multiple-keyhole MIMO channel in \cite{Levin2011} and a rich scattering MIMO channel in \cite{Rusek2013}. We consider a simple linear minimum mean square error (MMSE) receiver structure with successive interference cancellation (SIC) for joint information decoding. The proposed joint legacy-MRS pilot structure obviates changes on the legacy transmitter side. This enables an MRS node applicable in a backwards compatible manner.

The rest of this paper is organized as follows. Section \ref{sec:sysmod} introduces the channel and the signal models. Section \ref{sec:achievable rate_CSI} presents the ASR and LAR under full CSIR. In Section~\ref{sec:achievable rate_est}, we propose a pilot structure for joint channel estimation of the considered system and then study the impact of channel estimation errors. Simulation results and conclusions are provided in Section \ref{sec:Sim} and \ref{sec:Con}.

\vspace{-5pt}

\section{System model} \label{sec:sysmod}

\begin{figure}[!t]
\centering
\includegraphics[width=0.7\columnwidth]{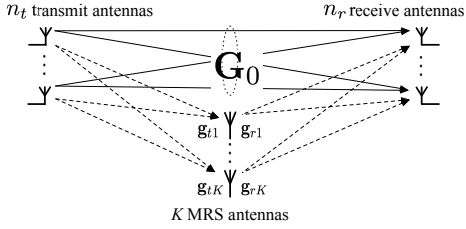}
\caption{An example of ambient MRS MIMO systems.}
\label{Fig:Channel}\vspace{-6pt}
\end{figure}
Consider an ambient MRS MIMO system with an $n_t$-antenna transmitter (TX), an $n_r$-antenna receiver (RX), and a $K$-antenna MRS node shown in Fig. \ref{Fig:Channel}. The MRS antennas are located in an array with no antenna element coupling problem. Otherwise, coupling would contribute to the channel uncertainty. The MRS is synchronised to the legacy system. The signal transmitted from the TX to the RX propagates through the direct links and paths passing through the MRS which modulates the signals.

Unless otherwise indicated, we adopt the following main assumptions: \begin{inparaenum}[1)~]
\item{the transmitted symbols of the legacy system are independently and identically distributed (i.i.d) standard circularly symmetric complex Gaussian (CSCG) distributed;}
\item{the CSI is available at the RX only;}
\item{the channels are block fading, and the channel coherence time is much longer than the frame duration. When channel estimation is considered, the training-based channel estimation and data transmission are assumed to be completed within the coherence time;}
\item{the channel vectors are i.i.d CSCG distributed, and are independent across the antennas of both systems.}
\end{inparaenum}
The assumptions have been considered in literature, e.g. \cite{Loyka2009, Levin2011} among others. The main notations are provided as footnote\footnote{\emph{Notations:} small and capital bold letters denote vectors and matrices, respectively; $\mathbf{A}^{\dagger}$ and $\mathbf{A}^T$ denote the Hermitian transpose and transpose of matrix $\mathbf{A}$; $\mathcal{CN}(\mathbf{0}, \mathbf{B})$ denotes the CSCG distribution with zero mean and covariance matrix $\mathbf{B}$; $\tr(\mathbf{X})$ and $\det(\mathbf{X})$ are the trace and determinant of a matrix $\mathbf{X}$; $\otimes$ is the Kronecker product operator; $\mathbf{I}_{n}$ is an $n\times n$ identity matrix, and the subscript $n$ may be omitted for simplicity; $||\cdot||$ is the Frobenius norm for matrices and Euclidean norm for vectors; $\mathbb{E}\{X\}$ is the expectation of $X$.}.

\vspace{-8pt}

\subsection{Channel Model}
Following the assumptions, the complete channel matrix between the transmitter and the receiver in $\mathbb{C}^{n_r\times (K+1)n_t}$ reads
\vspace{-3pt}
\begin{equation} 
	\mathbf{G}=\begin{bmatrix} 
	\mathbf{G}_0 &\mathbf{G}_1 & \cdots & \mathbf{G}_K 		\end{bmatrix}, \normalsize \label{composite_channel}
\end{equation} 
where $\mathbf{G}_0\in\mathbb{C}^{n_r\times n_t}$ denotes the channel matrix of the direct links. The channel matrix passing through the $k$\textsuperscript{th} MRS antenna $\mathbf{G}_k\in\mathbb{C}^{n_r\times n_t}$ for $ k = 1, \cdots, K$ is defined as
\vspace{-3pt}
\begin{equation}
	\mathbf{G}_k=\alpha\mathbf{g}_{rk}\mathbf{g}_{tk}^{\dagger}\,,
	\label{mrs_channel}
\end{equation}\vspace{-3pt}
where $\alpha\in\mathbb{C}$ is a constant scale factor\footnote{$\alpha$ could be used to model the impact of pathloss and absorption at the MRS antennas. The model could easily be extended to allow antenna-specific scaling factors.}, $\mathbf{g}_{tk}\in\mathbb{C}^{n_t}$ denotes the complex channel gains between transmit antennas and the $k$\textsuperscript{th} MRS antenna, and $\mathbf{g}_{rk}\in\mathbb{C}^{n_r}$ is the channel vector between the $k$\textsuperscript{th} MRS antenna and the receiver antennas. 

In this work, the channels passing through the MRS antenna $k$ resemble a MIMO keyhole channel \cite{Levin2011}. In our case, however, the legacy signals are additionally modulated by the MRS. If $\mathbf{G}_0=\mathbf{0}$, our general channel model coincides with the MIMO backscatter channels in \cite{He2015a,Boyer2014}.

\vspace{-5pt}

\subsection{Signal Model}
Consider a Wyner polyphase coding (WPC) modulation scheme at the MRS antennas, which has some important properties, such as periodic orthogonality, and the constant amplitude zero auto-correlation property \cite{Chu1972,Wyner1966}. The equivalent input over the considered MIMO channels in \eqref{composite_channel} is denoted by $\boldsymbol{\psi}$ (within one symbol) with
\begin{equation} \label{eq:def_psi}
    \boldsymbol{\psi} = 
    \left[\begin{matrix} 1 & \mathbf{x}_1 \end{matrix} \right]^T
     \otimes \mathbf{x}_0 
    =
    \left[\begin{matrix}
    \mathbf{x}_0 &
    \mathbf{x}_0 x_{11} &
    \cdots &
    \mathbf{x}_0 x_{1K}\end{matrix}\right]^T\,,
\end{equation}
where $\mathbf{x}_{0}\sim\mathcal{CN}(\mathbf{0}, \rho_d\mathbf{I}_{n_{t}})$ is the channel input vector of the transmitter at a given time instant. Here, $\rho_d$ denotes the data symbol power per antenna. We denote the reflection coefficients of the MRS antennas as $\mathbf{x}_1 \in\mathbb{C}^{K}$, where WPC is applied such that $\mathbb{E}\{\mathbf{x}_1\mathbf{x}^{\dagger}_1\}=\mathbf{I}_K$, $|x_{1k}|^2=1$, and $\mathbb{E}\{x_{1k}\}=0$ for each $k=1,\cdots, K$ \cite{Wyner1966}. Since each element of $\mathbf{x}_1$ is unitary, each element of $\boldsymbol{\psi}$ has the same distribution as the channel input, i.e., $\boldsymbol{\psi}\sim\mathcal{CN}(\mathbf{0}, \rho_d\mathbf{I}_{(K+1)n_{t}})$. The resulting discrete time signals for the considered channels Eq.~\eqref{composite_channel} within one frame of $N$ discrete samples reads
\begin{equation}
	\mathbf{Y} = \sqrt{\beta_K} \mathbf{G}\boldsymbol{\Psi} +\mathbf{Z}\, ,
	 \label{eq:Signal_modelY}
\end{equation}
where the $n_r\times N$ matrix $\mathbf{Y}$ denotes the received signals, the $n_t(K+1)\times N$ matrix  $\boldsymbol{\Psi}$ is the equivalent input whose columns $\boldsymbol{\psi}_{i}, \: \forall i=1,\cdots, N$ are given in Eq. \eqref{eq:def_psi}, the entries of the $n_r\times N$ noise matrix $\mathbf{Z}$ are i.i.d. $\mathcal{CN}(0, \sigma^2)$, and the scaling factor is $\beta_K=1/\left(n_{r}(K|\alpha|^{2}+1)\right)$. For a physical channel, the total received signal energy cannot exceed the total transmit energy. Since the average impact of the channel on the energy is $\mathbb{E}\left\{ \| \mathbf{G}\|^2\right\}$, in-line with \cite{Loyka2009}, channel vectors $\mathbf{g}_{d k}$ in (\ref{mrs_channel}) with $d\in \{r,t\}, \;\forall k$, and the matrix $\mathbf{G}$ in (\ref{composite_channel}) are normalised so that 
\begin{equation}
    \mathbb{E}\left\{||\mathbf{g}_{d k}||^{2}\right\}/n_{d} = 1, \text{ and } \mathbb{E}\left\{||\mathbf{G}||^{2}\right\}/n_{t} = n_{r}(K|\alpha|^{2}+1). \label{eq:chan_norml}
\end{equation}
Hence, with the signal model in Eq. \eqref{eq:Signal_modelY}, the total average received SNR from all transmit antennas yields
\begin{equation}
	\gamma = (P_{t}/n_{t}) \mathbb{E}\left\{||\mathbf{G}||^{2}\right\} \beta_K /\sigma^{2}  = P_{t}/\sigma^{2}\,,
	 \label{eq:snr}
\end{equation}
where $P_{t}=n_t \rho_d$ denotes the total transmit power. Thus, $\gamma/n_{r}$ and $\gamma/n_{t}$ represent the SNR per receive antenna and per transmit antenna, respectively, for a $K$-antenna MRS node.

\vspace{-2pt}

\section{Achievable rate with Full CSI at the Receiver} \label{sec:achievable rate_CSI}

We investigate the achievable rate of the considered system in Eq.~\eqref{eq:Signal_modelY} assuming that only the full CSIR is available, and the TX knows the channel distribution without having the instantaneous CSI. The Gaussian vector $\boldsymbol{\psi}$ defined in Eq.~\eqref{eq:def_psi} can be equivalently written as $\boldsymbol{\psi}=\sqrt{\rho_{d}}\mathbf{u}$ with $\mathbf{u}\sim\mathcal{CN}(\mathbf{0},\mathbf{I}_{n_t(K+1)})$ such that the received symbol vector reads
\begin{equation}
	\mathbf{y} = \sqrt{\rho_d\beta_K}{\mathbf{G}}\mathbf{u}+ \mathbf{z}.
\label{equivalentsystem}
\end{equation}

\begin{proposition}\label{sumrate}
The ASR of the system\footnote{Remark: The legacy system can achieve this achievable rate alone if polyphase-coded $\mathbf{x}_1$ is a known pseudo-random sequence at the receiver. That is, the MRS acts as a passive relay to assist the legacy.} in Eq.~\eqref{equivalentsystem} is 
\begin{equation*}
    \mathcal{R} = \log_2\det\left(\mathbf{I}+ \gamma \beta_{K} \mathbf{G}\mathbf{G}^{\dagger}/n_{t}\right).
\end{equation*}
\end{proposition} 

\begin{IEEEproof}
The channel model in Eq.~\eqref{equivalentsystem} corresponds to the standard MIMO system with zero mean circularly symmetric normal input $\mathbf{u}$ and channel matrix $ \sqrt{\rho_d\beta_K}{\mathbf{G}}$. Hence, the assertion follows from the results in \cite{Foschini1996}.
\end{IEEEproof}

\begin{proposition}\label{propos}
Let $\mathbf{F}=\gamma_{0}\mathbf{G}_0\mathbf{G}_0^{\dagger}$, and $\boldsymbol{\Delta}=\gamma_{0}\sum_{k=1}^K \mathbf{G}_k\mathbf{G}_k^{\dagger}$ with $\gamma_{0} = \gamma\beta_K/{n_{t}}$. Then, 
\begin{equation}\nonumber
\log_2\det\left(\mathbf{I}+\mathbf{F}\right) \le \log_2\det\left(\mathbf{I}+\mathbf{F}+\boldsymbol{\Delta}\right),
\end{equation}
\end{proposition}
with equality if and only if eigenspace of $\mathbf{F}$ is a subspace of the null space of $\boldsymbol{\Delta}$. 
\begin{IEEEproof}
By definition, $\mathbf{F}$ is a positive-semidefinite (PSD) Hermitian matrix with eigenvalues $\lambda_i(\mathbf{F})$, $\forall i=1,\dots, n_t$, and $\boldsymbol{\Delta}$ is a PSD Hermitian matrix. Thus, it is enough to show that the eigenvalues of a Hermitian PSD $\mathbf{F}$ are strictly less than the eigenvalues of $\mathbf{F}+\mathbf{\Delta}$. The Corollary of Weyl's theorem \cite[Corollary 4.3.3]{Horn2013} states that the eigenvalues fulfil
\begin{equation}\label{eq:weyl-eigen}
\lambda_i(\mathbf{F}+\boldsymbol{\Delta})\geq \lambda_{i}(\mathbf{F})
\end{equation}
with equality for some $i$ if and only if $\boldsymbol{\Delta}$ is singular and $\exists \mathbf{x}\neq \mathbf{0}$ such that $\mathbf{F} \mathbf{x}=\lambda_i(\mathbf{F})\mathbf{x}$, $\boldsymbol{\Delta}\, \mathbf{x}=\mathbf{0}$, and $(\mathbf{F}+\mathbf{\Delta}) \mathbf{x}=\lambda_i(\mathbf{F}+\boldsymbol{\Delta})\mathbf{x}$. Since $\log_2\det (\cdot)$ is equal to matrix trace, the achievable rate is increased in case only one of the eigenvalues satisfies  Eq.~\eqref{eq:weyl-eigen} with strict inequality. Conversely, $\mathbf{F}+\boldsymbol{\Delta}$ has the same trace as $\mathbf{F}$ if all eigenvectors of $\mathbf{F}$ are in the null space of $\boldsymbol{\Delta}$. The only PSD Hermitian matrix $\boldsymbol{\Delta}$ satisfying this condition is the all zero matrix. That is, equality can occur only if $\boldsymbol{\Delta}=\boldsymbol{0}$.
\end{IEEEproof}

\vspace{-6pt}

\subsection{Asymptotic Analysis}
We conduct asymptotic analysis of the considered system under complete CSIR. Note that, for an $n\times m$ matrix $\mathbf{A}$,
\begin{equation}\label{eq:trace-log-property}
\log_2\det(\mathbf{I}_{n}+\mathbf{A}\mathbf{A}^{\dagger})=\log_2\det(\mathbf{I}_{m}+\mathbf{A}^{\dagger}\mathbf{A}).
\end{equation}

\begin{corol}\label{achievable rate:n_r_infty}
Denote $\tilde{\gamma} = \gamma\beta_K$. As $n_r\to\infty$ with fixed $n_t$ and $K$, the LAR of the considered system is the ASR of a multiple-keyhole MIMO channel in \cite{Levin2011} and a rich scattering MIMO channel in \cite{Rusek2013}, i.e.\footnote{Corollary \ref{achievable rate:n_r_infty} indicates that the additional MRS antennas increase the overall system achievable rate due to the increased rank of the overall channel matrix.}
\begin{IEEEeqnarray*}{ll}
    \mathcal{R}_{\mathsf{LAR}}^{n_r} &=n_t\log_2\left(1+\tilde{\gamma}\right) + \sum_{k=1}^K\log_2\left(1+|\alpha|^2||\mathbf{g}_{tk}||^2 \, \tilde{\gamma}\right).
\end{IEEEeqnarray*}
\end{corol}

\begin{IEEEproof}
Using Proposition~\ref{sumrate}, Eq.~\eqref{eq:chan_norml}, and Eq.~\eqref{eq:trace-log-property}, as ${n_r\to\infty}$ we have
\begin{equation*}
\frac{1}{n_r}\mathbf{G}^{\dagger}\mathbf{G}\to \begin{bsmallmatrix}
 \mathbf{I}_{n_{t}} & \mathbf{0}                            & \cdots &\mathbf{0}\\
\mathbf{0}         & |\alpha|^2\mathbf{g}_{t1}^{\dagger}\mathbf{g}_{t1} & \cdots &\mathbf{0}\\
\vdots             & \vdots                                & \ddots &\vdots\\
\mathbf{0}        & \mathbf{0}                            & \cdots & |\alpha|^2\mathbf{g}_{tK}^{\dagger}\mathbf{g}_{tK}
 \end{bsmallmatrix}
\end{equation*}
\normalsize
with eigenvalues 
\begin{equation*}
 \underbrace{1,  \cdots, 1}_{n_t},  \underbrace{|\alpha|^2 ||\mathbf{g}_{t1}||^2,  \cdots, |\alpha|^2 ||\mathbf{g}_{tK}||^2}_{K}, \underbrace{0,  \cdots,  0}_{(n_t-1)K}.
\end{equation*}
We obtain the result from Proposition \ref{sumrate}.
\end{IEEEproof}

\begin{corol}\label{achievable rate:n_t_infty2}
Let $\tilde{\gamma} = \gamma\beta_K$. As $n_t\to\infty$ with fixed $n_r\geq K$, the achievable rate of the considered system reads
\begin{equation*}
    \mathcal{R}_{\mathsf{LAR}}^{n_t} =(n_r-K) \log_2\left(1+\tilde{\gamma}\right) + \sum_{k=1}^K\log_2\left(1+|\alpha|^2||\mathbf{g}_{rk}||^2 \tilde{\gamma}\right).
\end{equation*}
\end{corol}
\begin{IEEEproof}
Following the same line of reasoning as the Corollary \ref{achievable rate:n_r_infty}, as $n_t\to\infty$, we have
\small
\begin{equation*}
\frac{1}{n_t}\mathbf{G} \mathbf{G}^{\dagger}\to 
 \mathbf{I}_{n_r} + \sum_{k=1}^K |\alpha|^2\mathbf{g}_{rk}\mathbf{g}_{rk}^{\dagger}.
\end{equation*}
\normalsize
Vectors $\mathbf{g}_{rk}, \forall k$ become asymptotically orthogonal as $n_r$ grows with fixed $K$. Hence, the eigenvalues of $\mathbf{G} \mathbf{G}^{\dagger}/n_t$ read
\begin{equation*}
 \underbrace{1+|a_1|^2 ||\mathbf{g}_{r1}||^2,\,  \cdots,\, 1+|\alpha|^2 ||\mathbf{g}_{rK}||^2}_{K}, \underbrace{1,  \cdots,  1}_{n_r-K}.
\end{equation*}
The result follows.
\end{IEEEproof}

\vspace{-6pt}
\subsection{Achievable rate of the MRS when $K=1$}
Obtaining the exact achievable rate for a constant envelope MIMO system is problematic. Hence, we focus on the case $K=1$ in order to gain an insight. The received signal reads
\begin{equation}
	\mathbf{y} =  \sqrt{\beta_1}\mathbf{G}
\left(\left[\begin{matrix} 1 & x_1 \end{matrix} \right]^T
 \otimes \mathbf{x}_0\right) + \mathbf{z}.
	\label{eq:receivedSig_K1}
\end{equation}
where $\mathbf{G} = \begin{bmatrix}\mathbf{G}_0 & \mathbf{G}_1\end{bmatrix}$ is defined in Eq. \eqref{composite_channel}, and $\beta_{1}$ is given by $\beta_K$ defined in (\ref{eq:Signal_modelY}) when $K=1$. Assume that the receiver would be able to decode $\mathbf{x}_0$ first. One approach to obtain $\mathbf{x}_0$ first is to adopt a linear MMSE receiver with SIC acting upon the matrix $\mathbf{G}$. The MMSE-SIC receiver successively decodes the strongest stream of the legacy system first, and then after removing the stream, it decodes the remaining strongest stream. Without loss of generality, we assume that the strength of the flows of the legacy system are sorted in a decreasing order. Conditioned on the channel matrices $\mathbf{G}_0$ ($x_1$ is unknown, and so is $\mathbf{G}_1 x_1$), the SINR for the currently decoded flow, i.e. the $i$th flow, of the legacy system can be expressed as \cite{Tse2005} 
\begin{equation*}
     \gamma_{0,i}=\left(\left[\left(\mathbf{I}+\beta_1\gamma\mathbf{G}_{>i}^\dagger\mathbf{G}_{>i}/n_t\right)^{-1}\right]_{11}\right)^{-1}-1, \; \forall i=1,\cdots, n_t,
\end{equation*}
where $\mathbf{G}_{>i}\stackrel{def}{=}\begin{bmatrix} 
	\mathbf{G}_0[>i] &\mathbf{G}_1 \end{bmatrix}$ is the matrix whose columns associated to the decoded streams of the legacy user have been removed. Hence, the achievable rate of the legacy system is 
\begin{equation}
    \mathcal{R}_{0} =  \sum_{i=1}^{n_t} \log_2(1+\gamma_{0,i}).
    \label{eq:r_0}
\end{equation}
Once $\mathbf{x}_0$ has been decoded, we can treat it as a known part and cancel $\mathbf{G}_0\mathbf{x}_0$ from the received signal in Eq. (\ref{eq:receivedSig_K1}). The receiver for decoding $x_1$ is operating based on the measurement $\mathbf{y}_1=\sqrt{\beta_1}(\mathbf{G}_1\mathbf{x}_0)x_1+\mathbf{z}$. The SNR for the MRS flow using a matched filter receiver (in a single user case, the SNR of a matched filter is identical to the one of an MMSE \cite{Tse2005}) yields 
\begin{equation}
    \gamma_{1}(\mathbf{x}_0)= \beta_{1} \mathbf{x}_0^\dagger\mathbf{G}_1^\dagger\mathbf{G}_1\mathbf{x}_0/\left(n_{t}\sigma^2\right).
\end{equation}
Hence, the achievable rate using the Gaussian codebook reads\footnote{The dependence of $\gamma_1$ on $\mathbf{x}_0$ is dropped from the notation for simplicity.}
\begin{equation}
    \mathcal{R}_{\mathsf{GC}}(\gamma_{1}) \leq \log_2\left(1 + \gamma_{1}\right).
\label{eq:cmbsG}
\end{equation}
For WPC,  the channel achievable rate is given in \cite{Wyner1966} by  
\begin{equation}
    \mathcal{R}_{\mathsf{WPC}}(\gamma_{1}) = -\int_{0}^{\infty} f(u,\gamma_{1}) \ln\left(\frac{f(u,\gamma_{1})}{u}\right) \text{d} u + \ln\left(\frac{2\gamma_{1}}{e}\right),
\label{eq:cmbspoly}
\end{equation}
where $
	f(u,\gamma_{1}) = 2u \gamma_{1} e^{-\gamma_{1}(1+u^{2})}J_{0}(2u \gamma_{1}),
$ and $J_{\nu}(x)$ is the modified Bessel function of $\nu$\textsuperscript{th} order. If $\gamma_{1}$ is close to zero, $R_{\text{WPC}}(\gamma_{1})$ can be approximated to 
$
R_{\text{WPC}}(\gamma_{1}) \approx \gamma_{1}
$
\cite{Wyner1966}. We note that if $x_1$ is known with $\mathbf{G}_1$ at the receiver, the achievable sum rate coincides with Proposition \ref{sumrate}, and that is the maximal rate that the legacy user could achieve using MMSE-SIC receiver acting on the channel matrix $\mathbf{G}_0+\mathbf{G}_1x_1$ \cite{Tse2005}.

\begin{figure}[!t]
\centering
\includegraphics[width=0.8\columnwidth]{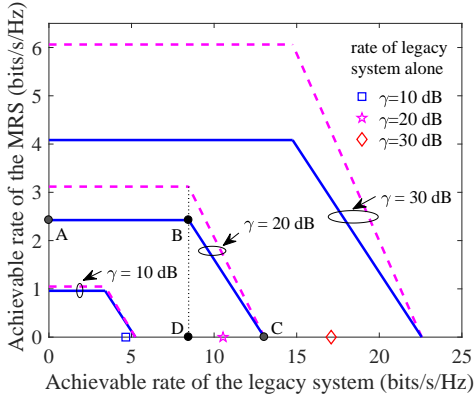} 
\caption{Achievable rate regions by the legacy system and the single-antenna MRS with $n_t=2, n_{r}=4$, and $\gamma=\left\{10, 20, 30\right\}$dB. The solid and dashed curves show the regions where the MRS system employs the WPC scheme and the Gaussian codebook, respectively.}
\label{fig:rateRegion}
\end{figure}

Provided that the legacy system uses Gaussian codebook, Fig. \ref{fig:rateRegion} depicts the achievable rate regions for a $2\times 4$ legacy system and a single-antenna MRS system employing the considered MMSE-SIC decoding scheme. Each plot is averaged over $2\times 10^{5}$ realisations. Select $\gamma = 20$dB  and the WPC scheme. Vertex A, denoting $\left\{0,\mathbb{E}\left[\mathcal{R}_{\mathsf{WPC}}(\gamma_{1})\right]\right\}$, shows the achievable rate of the MRS when the legacy transmitter transmits signals containing no information. Vertex B represents $\left\{\mathbb{E}\left[\mathcal{R}_{0}\right],\; \mathbb{E}\left[\mathcal{R}_{\mathsf{WPC}}(\gamma_{1})\right]\right\}$. Hence, segment AB denotes that the legacy system is not assisted by the MRS. Vertex C, denoting $\left\{\mathbb{E}\left[\mathcal{R}\right],\; 0\right\}$, shows the ASR given by Proposition \ref{sumrate}, and is achievable by the legacy system alone when $\mathbf{x}_{1}$ is known and a joint detection scheme is used at the receiver. Segment BC can be achieved by the legacy and the MRS systems through a time-duplexing scheme. Point D shows the achievable rate by the legacy system when $\mathbf{x}_{1}$ is unknown by the legacy system and treated as noise. We also plot the achievable rate of the legacy system alone to show the achievable rate improvement when the MRS assists the legacy system\footnote{As K=0, the channel normalization coefficient $\beta_0=1/n_r$.}.

\vspace{-6pt}
\section{Achievable rate with Channel Estimation}
\label{sec:achievable rate_est}
We consider an LS channel estimator to study the impact of channel estimation errors on the performance. In addition, we propose a pilot structure based on Hadamard matrix to jointly estimate the direct and re-scattered links.

\vspace{-5pt}
\subsection{Joint channel estimation} \label{sec:channEst}
Suppose that the data transmission with a total coherence interval of length $N$ consists of two phases: a training phase and a data transmission phase \cite{Hassibi2003}. In the training phase, a pilot sequence is transmitted to the receiver to estimate the channel matrix $\mathbf{G}$ in Eq.~\eqref{composite_channel} using acquired $N_p$ digital samples. Since our aim is to investigate the impact of channel estimation on the system achievable rate, we first describe the pilot sequence design. The LS estimation error and its impact on the achievable rate are studied for the designed pilots.

Since we assume that the MRS is synchronised to the legacy transmitter, this casts the current pilot design problem to the one for a multi-relay system in \cite{Nasir2013}. Hence, one technique is to generate the pilot sequences using Hadamard matrices, which has recursive generation property. Let $\mathbf{H}_{n}$ denote a $2^n \times 2^n$ dimensional Hadamard matrix. Then, starting from ${\mathbf{H}_{1}=\begin{bsmallmatrix}1 & 1 \\ 1 & -1\end{bsmallmatrix}}$, a $2^n$-dimensional matrix can be generated from the recursion $	\mathbf{H}_{n}= \mathbf{H}_1\otimes \mathbf{H}_{n-1}$. We assume that the legacy system uses orthogonal pilots, and the MRS uses rows of $\mathbf{H}_{n}$, such that the resulting joint pilot becomes orthogonal. 

Let $\mathbf{X}_{0p}\in\mathbb{C}^{n_t\times m_0}$ and $\mathbf{X}_{1p}\in\mathbb{C}^{(K+1)\times m_1}$ be the orthogonal sequences of the legacy system and the MRS with length of $m_0$ and $m_1$, respectively, yielding $N_{p} = m_{0} m_{1}$. The two sequences have the properties $\mbox{E}\{\mathbf{X}_{0p}\mathbf{X}_{0p}^{\dagger}\} = m_0 \rho_p \mathbf{I}_{n_t}$ and $\mbox{E}\{\mathbf{X}_{1p}\mathbf{X}_{1p}^{\dagger}\}=m_1\mathbf{I}_{K+1}$, where $\rho_p$ denotes the power of a pilot and could be different from $\rho_d$ the data symbol power\footnote{See \cite{Hassibi2003} for the impact of $\rho_p$ and $\rho_d$ on the achievable rate.}. All elements of the first row of $\mathbf{X}_{1p}$ are equal to $1$, which correspond to the paths $\mathbf{G}_0$ with constant non-controllable channel gains. Hence, $\mathbf{X}_{1p}$ could be selected to correspond to the first $K+1$ rows of the Hadamard matrix so that the composite pilot, with dimension $n_t(K+1)\times(m_0 m_1)$, $\boldsymbol{\Psi}_{p}=\mathbf{X}_{1p}\otimes \mathbf{X}_{0p}$ is orthogonal, and $\boldsymbol{\Psi}_{p}\boldsymbol{\Psi}_{p}^{\dagger}= m_0 m_1 \rho_p \mathbf{I}_{n_t(K+1)}$. The composite pilots $\boldsymbol{\Psi}_{p}$ require the MRS pilot symbol length to be $m_0$ legacy symbols. Hence, the training phase can be realised by transmitting the same pilot sequence of the legacy systems $K+1$ times so that during the transmission of one repetition of the legacy pilot sequence, the MRS pilot remains constant but it is changed from repetition to another.. Following this strategy and using the signal model in Eq. \eqref{eq:Signal_modelY}, the received $n_r\times m_0 m_1$ pilots matrix $\mathbf{Y}_p$ yields $
	\mathbf{Y}_{p} = \sqrt{\beta_K}\mathbf{G}\boldsymbol{\Psi}_{p} +\mathbf{Z}_{p}$, where the entries of the noise matrix $\mathbf{Z}_{p}$ are i.i.d. standard complex Gaussian random variables. Knowing the transmitted pilots $\boldsymbol{\Psi}_p$ and the received pilots matrix $\mathbf{Y}_p$, the LS estimate of the (normalized) channel matrix reads
\begin{equation}
    \sqrt{\beta_K}\hat{\mathbf{G}} =  \mathbf{Y}_p\boldsymbol{\Psi}_p^{\dagger}(\boldsymbol{\Psi}_{p}\boldsymbol{\Psi}_p^{\dagger})^{-1} = \mathbf{Y}_p\boldsymbol{\Psi}_p^{\dagger}/\left(m_0m_1\rho_p\right).
\end{equation}
The (normalized) channel estimation error is
\small
\begin{equation}\label{eq:channel-estimation-error}
	\sqrt{\beta_K}\tilde{\mathbf{G}}=  \mathbf{Z}_{p}\boldsymbol{\Psi}_p^{\dagger}(\boldsymbol{\Psi}_{p}\boldsymbol{\Psi}_p^{\dagger})^{-1} = \mathbf{Z}_{p}\boldsymbol{\Psi}_p^{\dagger}/\left(m_0m_1\rho_p\right).
\end{equation}
\normalsize

\vspace{-7pt}

\subsection{Achievable Rate with Channel Estimation Error} \label{sec:c_{Est}}
Considering channel estimation errors, the capacity of a MIMO system is not known \cite{Hassibi2003,Yoo2006}. However, the channel uncertainty can be treated as noise when the channel estimate $\hat{\mathbf{G}}$ is available at the receiver. As the considered estimator is unbiased and the estimation error in Eq.~\eqref{eq:channel-estimation-error} is Gaussian, the received data signal vector of a data symbol transmission reads
\begin{equation}
\begin{split} 
	\mathbf{y} = \sqrt{\beta_K}{\hat{\mathbf{G}}} \boldsymbol{\psi} + \sqrt{\beta_K}\tilde{\mathbf{G}}\boldsymbol{\psi} + \mathbf{z}.
\end{split} 
\end{equation}
The covariance matrix of $\sqrt{\beta_k}\tilde{\mathbf{G}}\boldsymbol{\psi}$ is given by
\begin{equation}
\mathbf{R}_{\mathbf{\tilde{\mathbf{G}}\boldsymbol{\psi}}}=\left(m_0m_1\rho_p\right)^{-2}\mathbb{E}\left\{\mathbf{Z} \boldsymbol{\Psi}_p^{\dagger}\boldsymbol{\psi} \boldsymbol{\psi}^{\dagger}\boldsymbol{\Psi}_p\mathbf{Z}^{\dagger}\right\}.
\end{equation}

As the pilots are known at the receiver, we obtain
\begin{equation}
	\mathbf{R}_{\mathbf{\tilde{\mathbf{G}}\boldsymbol{\psi}}}=\frac{\rho_d \sigma^2\mathbf{I}_{n_r}}{(m_0m_1\rho_p)^2}\tr\left(\boldsymbol{\Psi}_p \boldsymbol{\Psi}_p^{\dagger}\right) = \frac{n_t(K+1)\rho_d}{\rho_p m_0m_1} \sigma^2 \mathbf{I}_{n_r}.
	\label{eq:estimationerror}
\end{equation}
Consequently, a lower bound of the ASR is obtained as \cite{Yoo2006} 
\begin{equation}
    \mathcal{R}_{\mathsf{est}}  \geq \frac{N-N_{p}}{N}\log_2\det\left(\mathbf{I}+ \mathbf{W}_K^{-1}\hat{\mathbf{G}}\hat{\mathbf{G}}^{\dagger}\gamma \beta_{K} \sigma^2\right),
\end{equation}
where $\mathbf{W}_K = \left(\sigma^2\mathbf{I}_{n_r} + \mathbf{R}_{\mathbf{\tilde{\boldsymbol{G}}\boldsymbol{\psi}}}\right)$ and $\mathbf{R}_{\mathbf{\tilde{\boldsymbol{G}}\boldsymbol{\psi}}}$ is given in (\ref{eq:estimationerror}).

\section{Simulation results}\label{sec:Sim}
We present the simulation results of ASR of the considered system under perfect and imperfect CSIR. Particularly, we show the impacts of the MRS system on the achievable rate of the overall system, and the achievable rate of the MRS system when $K=1$. The simulation results are averaged over $2\times 10^{5}$ realisations, where for each realisation the pilot and data symbol powers are the same, $\rho_{d}=\rho_{p}$, and the amplitude of the attenuation shift factor of the MRS is $|\alpha| = -3$ dB.

\begin{figure}[!t]
\centering
\includegraphics[width=0.8\columnwidth]{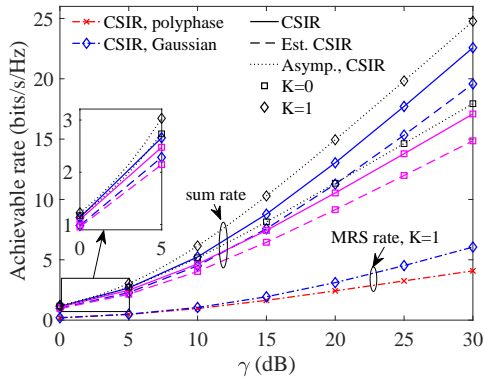} 
\caption{Sum rate with perfect and imperfect CSIR for $K = 0,1$, $n_t=2$, $n_{r}=4$, $\rho_{p}=\rho_{d}$, $m_0m_1=128$ and $N=1000$.}
\label{fig:C_est_rhod}
\end{figure}

Fig. \ref{fig:C_est_rhod} shows the achievable rate by the whole system as a function of $\gamma$ for $K=\{0,\; 1\}$, $n_t=2$ and $n_{r}=4$. The achievable rate by the MRS system alone ($K=1$) is plotted when the MRS employs the WPC code book. For the MRS also an upper-bound for the achievable rate with Gaussian coding schemes is shown. The solid and dashed curves depict the achievable sum rate with perfect and imperfect CSI at the receiver, respectively. The asymptotic ASR as $n_r\to \infty$ is plotted using dotted curves. The simulation results confirm Proposition \ref{propos}, i.e., under perfect CSIR, the ASR always exceeds that the legacy system could achieve alone. Adding an MRS node is beneficial even in low SNR, though the rate gain is more visible in the high SNR domain. Furthermore, in low SNR domain, LAR and ASR are close to each other. The channel estimation errors will reduce the ASR of the system.

\section{Conclusion}\label{sec:Con}
This work studied a system consisting of a legacy MIMO link and a multi-antenna MRS node. We showed that the ASR of such a system is larger than that which the legacy MIMO system could achieve alone. Asymptotically in rich scattering case (either $n_t$ or $n_r$ approach to infinity), the LAR becomes the sum of the LAR of the multiple keyhole MIMO channel and the LAR of the rich scattering MIMO channel. We proposed to use MMSE-SIC receiver for joint decoding of the legacy and MRS symbols at the receiver. For the channel estimation, we proposed a simple Hadamard matrix based pilot structure that allows the legacy MIMO transmitter to use its pilot sequences without any modifications. We believe that our work opens the possibility to enrich existing communication systems with a layer of ultra-low-power MRS nodes.

%
%

\ifCLASSOPTIONcaptionsoff
  \newpage
\fi

\bibliographystyle{IEEEtran}

\begin{thebibliography}{10}

\bibitem{Boyer2014}
C.~Boyer and S.~Roy, ``Backscatter communication and {RFID}: Coding, energy, and {MIMO} analysis,'' \emph{IEEE Trans. Commun.}, vol.~62, no.~3, pp. 770-785, March 2014.

\bibitem{He2015a}
C.~He, Z.~Wang, and V.~Leung, ``Unitary query for the {$M\times L\times N$} {MIMO} backscatter {RFID} channel,'' \emph{IEEE Trans. Wireless Commun.}, vol.~14, no.~5, pp. 2613-2625, May 2015.

\bibitem{Kimionis2014}
J.~Kimionis, A.~Bletsas, and J.~Sahalos, ``Increased range bistatic scatter
radio,'' \emph{IEEE Trans. Commun.}, vol.~62, no.~3, pp. 1091-1104, Mar. 2014.


\bibitem{Bharadia2015}
D.~Bharadia, K.~R. Joshi, M.~Kotaru, and S.~Katti, ``{BackFi}: High throughput {WiFi} backscatter,'' \emph{ACM SIGCOMM Comput. Commun. Rev.}, vol.~45, no.~5, pp. 283-296, Aug. 2015.

\bibitem{Wu2016}
Q. Wu, M. Tao, D. W. Kwan Ng, W. Chen and R. Schober,
"Energy-efficient resource allocation for wireless powered communication networks",  \emph{IEEE Trans. Wireless Commun.}, vol.~15, no.~3, pp. 2312-2327, Mar. 2016.

\bibitem{Wu2016a}
Q. Wu, W. Chen, D. W. Kwan Ng, J. Li and R. Schober,
"User-centric energy efficiency maximization for wireless powered communications",  \emph{IEEE Trans. Wireless Commun.}, vol.~15, no.~10, pp. 6898-6912, Oct. 2016.

\bibitem{Wang2016}
G.~Wang, F.~Gao, R.~Fan, and C.~Tellambura,
"Ambient backscatter communication systems: Detection and performance analysis",  \emph{IEEE Trans. Commun.}, vol.~64, no.~11, pp. 4836-4846, Nov. 2016.

\bibitem{Qian2017}
J.~Qian, F.~Gao, G.~Wang, S.~Jin, and H.~Zhu,
"Noncoherent detections for ambient backscatter system",  \emph{IEEE Trans. Wiless Commun.}, vol.~16, no.~3, pp. 1412-1422, Mar. 2017.

\bibitem{Darsena2017}
D.~Darsena and G.~Gelli and F.~Verde,
"Modeling and performance analysis of wireless networks with ambient backscatter devices", \emph{IEEE Trans. Commun.}, 2017, in press.

\bibitem{Levin2011}
G.~Levin and S.~Loyka, ``From multi-keyholes to measure of correlation and power imbalance in {MIMO} channels: Outage achievable rate analysis,'' \emph{IEEE Trans. Inf. Theory}, vol.~57, no.~6, pp. 3515-3529, Jun. 2011.

\bibitem{Rusek2013}
F.~Rusek, D.~Persson, B.~Lau, E.~Larsson, T.~Marzetta, O.~Edfors, and F.~Tufvesson, ``Scaling up {MIMO}: Opportunities and challenges with very large arrays,'' \emph{IEEE Signal Process. Mag.}, vol.~30, no.~1, pp. 40-60, Jan. 2013.

\bibitem{Loyka2009}
S.~Loyka and G.~Levin, ``On physically-based normalization of {MIMO} channel matrices,'' \emph{IEEE Trans. Wireless Commun.}, vol.~8, no.~3, pp. 1107-1112, Mar. 2009.


\bibitem{Chu1972}
D.~Chu, ``Polyphase codes with good periodic correlation properties,'' \emph{IEEE Trans. Inf. Theory}, vol.~18, no.~4, pp. 531-532, Jul. 1972.

\bibitem{Wyner1966}
A.~D. Wyner, ``Bounds on communication with polyphase coding,'' \emph{Bell Syst. Tech. J.}, vol.~45, pp. 523-559, 1966.

\bibitem{Foschini1996}
G.~J. Foschini, ``Layered space-time architecture for wireless communication in a fading environment when using multi-element antennas,'' \emph{Bell Labs Tech. J.}, vol.~2, pp. 41-59, 1996.

\bibitem{Horn2013}
R.~A. Horn and C.~R. Johnson, \emph{Matrix Analysis}, 2nd~ed. \relax Cambridge University Press, 2013.

\bibitem{Tse2005}
D.~Tse and P.~Viswanath, \emph{Fundamentals of Wireless Communication}. \relax Cambridge University Press, 2005.

\bibitem{Hassibi2003}
B.~Hassibi and B.~M. Hochwald, ``How much training is needed in multiple-antenna wireless links?'' \emph{IEEE Trans. Inf. Theory}, vol.~49, no.~4, pp. 951-963, Apr. 2003.

\bibitem{Nasir2013}
A.~A. Nasir, H.~Mehrpouyan, S.~Durrani, S.~D. Blostein, R.~A. Kennedy, and B.~Ottersten, ``Optimal training sequences for joint timing synchronization and channel estimation in distributed communication networks,'' \emph{IEEE Trans. Commun.}, vol.~61, no.~7, pp. 3002-3015, Jul. 2013.

\bibitem{Yoo2006}
T. Yoo and A. Goldsmith, ``Achievable rate and power allocation for fading {MIMO} channels with channel estimation error,'' \emph{IEEE Trans. Inf. Theory}, vol.~52, no.~5, pp. 2203-2214, May 2006.

\end{thebibliography}

\end{document}